\documentclass[aps,prx,twocolumn,preprintnumbers,amsmath,amssymb,superscriptaddress]{revtex4-1}
\usepackage{graphicx}
\usepackage{subfigure}
\usepackage{epsfig}
\usepackage{dcolumn}
\usepackage{bm}
\usepackage{ulem}
\usepackage{color}
\usepackage{multirow}
\usepackage{slashed}

\def\be{\begin{equation}}       \def\ee{\end{equation}}
\def\bea{\begin{eqnarray}}      \def\eea{\end{eqnarray}}
\def\ba{\begin{array}}
\def\ea{\end{array}}
\def\bnum{\begin{enumerate} }
\def\enum{\end{enumerate}}

\def\nn{\nonumber}

\def\=>{\Rightarrow}
\def\>{\rightarrow}

\def\eye2{Fathbb{I}}

\def\Eq#1{Eq.~(\ref{#1})}

\renewcommand{\>}{\rangle}

\begin{document}

\title{Universal properties of many-body quantum chaos at Gross-Neveu criticality}

\author{Shao-Kai Jian}
\affiliation{Institute for Advanced Study, Tsinghua University, Beijing 100084, China}
\author{Hong Yao}
\email{yaohong@tsinghua.edu.cn}
\affiliation{Institute for Advanced Study, Tsinghua University, Beijing 100084, China}
\affiliation{State Key Laboratory of Low Dimensional Quantum Physics, Tsinghua
University, Beijing 100084, China}

\begin{abstract}
Quantum chaos in many-body systems may be characterized by the Lyapunov exponent defined as the exponential growth rate of out-of-time-order correlators (OTOC). So far Lyaponov exponents around various quantum critical points (QCP) remain largely unexplored. Here, we investigate the Lyapunov exponent around QCPs of the Gross-Neveu (GN) model with $N$ flavors of Dirac fermions in (2+1) dimensions. Around the GN quantum phase transition between a Dirac semimetal and a gapped insulator breaking $Z_2$ symmetry (e.g., inversion symmetry of the honeycomb lattice), we find that the Lyaponov exponent $\lambda_L \approx 3.5 T/N$ at temperature $T$ and to the leading order of $1/N$ in the large-$N$ expansion. We also obtain the quantum scattering rate of an excitation with energy $\epsilon$, which is proportional to $\sqrt{\epsilon T}/N$ at low energy. We further discuss possible experimental relevances of the GN model in many-body systems.
\end{abstract}
\date{\today}
\maketitle

\section{Introduction}

Fathoming the dynamic properties of quantum many-body systems is among the central questions in modern condensed matter physics. Especially, understanding the physics of thermalization and quantum chaos precisely in an isolated quantum system \cite{deutsch1991, srednicki1994} remains challenging. Recently, it has attracted increasing attentions owing to the exciting advances made in both theories and experiments, including the progresses in solving black hole information paradoxes \cite{preskill2007, susskind2008} and the experimental realizations of nearly isolated quantum systems \cite{rigol2008, langen2013, greiner2016}. It is now understood that starting from a generic state with local or global perturbations, at an intermediate time scale before equilibrium which is usually referred to as scrambling time \cite{susskind2008},
the initial information about the state spreads across the whole system and cannot be restored via local measurements. The quantum chaos is closely related to the onset of scrambling, and measures how fast a state can scramble in a given physical system.

It was recently proposed that the quantum chaos may be characterized by a quantum version of Lyapunov exponent defined as the exponential growth rate the out-of-time-order correlator (OTOC) \cite{larkin1969, shenker2014,shenker2014b, kitaev2014}. The proposal has triggered a surge of interests in investigating OTOC in various physical models, ranging from continuous field theories \cite{polchinski2016b, stanford2016c, swingle2016a, ioffe2016, sachdev2017a, moessner2017, swingle2017a, sachdev2017b, berg2017, schmalian2017, vishwanath2018}, holographic theories \cite{ling2016, xian2017, wu2018, setty2018, blake2017, blake2017b}, to lattice models \cite{knap2017, lev2017, ueda2017, shen2017, fradkin2016, chen2016, swingle2017b, lu2017, chenyu2016, fan2017, you2017, huse2018}. Several protocols \cite{swingle2016b, moore2016, grover2016} of experimentally measuring OTOC are proposed and experiments have been done \cite{rey2017, zeng2017, wei2018}. Among all known systems with local interactions, black holes are shown to be the fastest scrambler \cite{susskind2008, stanford2016a} in nature with $\lambda_L=2\pi T$, where $T$ is the black hole temperature. In particular, significant progress has been made by studies of the Sachdev-Ye-Kitaev model \cite{kitaevtalk2015, sachdev1993, polchinski2016a, stanford2016b}, which is dual to a gravitational system and (almost) saturates the upper bound of Lyapunov exponent \cite{stanford2016a}.


Along the lines with AdS/CFT correspondence \cite{maldacena1998, witten1998, hartnoll2016}, a strongly-coupled many-body system having a gravity dual is expected to be highly chaotic.
Although generic many-body systems do not respect conformal symmetry in microscopic scales, critical points \cite{sachdevbook} separating distinct phases 
feature strong fluctuations and possess emergent conformal symmetry in low energy and long distance. Thus, besides some critical phases \cite{sachdev2017b, schmalian2017}, it is natural to suspect that many-body systems at critical points are fast scramblers in nature. Because temperature is the only energy scale in a critical theory, the Lyapunov exponent is expected to obey $\lambda_L\sim \kappa T$ at low temperature $T$, where $\kappa$ is a universal number, associated the universality class of QCP under consideration but irrespective of the microscopic details. It was shown recently that the (2+1)-dimensional bosonic $O(N)$ transition \cite{swingle2017a, justin2003} exhibits $\lambda_T \approx 3.2T/N$, at low temperature and to the leading order of $1/N$  at the large-$N$ limit \cite{swingle2017a}.



In this paper, we investigate universal properties of OTOC around fermionic quantum critical points involving fermionic degrees of freedom in (2+1) dimensions. Specifically, we study the Gross-Neveu (GN) model \cite{justin2003, nambu1961a, nambu1961b, gross1974}, which describes a $Z_2$ quantum phase transition in materials hosting massless Dirac fermions such as graphene and graphene-like materials \cite{firsov2005, geim2009}, by computing its relaxation and scrambling at finite temperature and at large-$N$ limit. The quantum scattering rate of Dirac fermion is given by the imaginary part of self-energy, equivalent to the inverse lifetime \cite{mirlin2011}. We find that in the low energy, the quantum scattering rate is proportional to $\sqrt{\epsilon T}/N$, where $\epsilon$ is the energy of excitations. Since $\sqrt{\epsilon}$ overwhelms the corresponding energy scale $\epsilon$ at low-energy limit, it implies the breakdown of quasiparticle picture, consistent with the renormalization group analysis \cite{justin2003, kivel1994} of the GN model. The behavior of the scattering rate in the low energy is similar to the case where Dirac fermions interact via long-range Coulomb interactions \cite{mirlin2011}. Moreover, we calculate the Lyapunov exponent $\lambda_T$ defined by OTOC:
\bea
	C(t) \sim \langle |\{\psi(t), \psi^{\dag}(0) \}|^2 \rangle \propto e^{\lambda_L t},
\eea
where $\psi$ is a schematic representative of fermionic operator, $\{, \}$ denotes the anti-commutator, and $\langle \rangle$ means the average over the thermal ensemble. We approximate the kernel function to the lowest nontrivial order in $1/N$ expansion. The calculation of Lyapunov exponent reduces to finding the largest eigenvalue of an integral \cite{stanford2016c}, which is done numerically by discretizing the integration and the kernel function. To the leading order of $1/N$, we find that $\lambda_L \approx 3.5T/N$ which is in agreement with the upper bound $2\pi T$. The universal coefficient $\kappa \approx 3.5/N$ decreases with increasing $N$, which is expected since quantum fluctuations should get weaker for larger $N$.

The rest of the paper is organized as follows: In Sec. II, we introduce the GN model and set up the notations for carrying out the $1/N$ expansion of the quantum scattering rate and the Lyapunov exponent. The polarization function at finite temperature is calculated in Sec. III. These functions have essential contributions to the quantum scattering rate and the Lyapunov exponent. In Sec. IV, we compute the quantum scattering rate of the Dirac fermions. The result shows that quasiparticle does not exist in the quantum critical region at finite temperature. We obtain the Bether-Saltpeter equation governing the exponentially growing part of OTOC in Sec. V. In Sec. VI, numerical calculations of Lyapunov exponent are carried out. We discuss the relation to Gross-Neveu-Yukawa model in Sec. VII. The appendices contain the detailed calculations of various quantities related to the GN model. We set $\hbar=k_B=1$ for simplicity in this paper.

\begin{figure}[t]
\includegraphics[width=6cm]{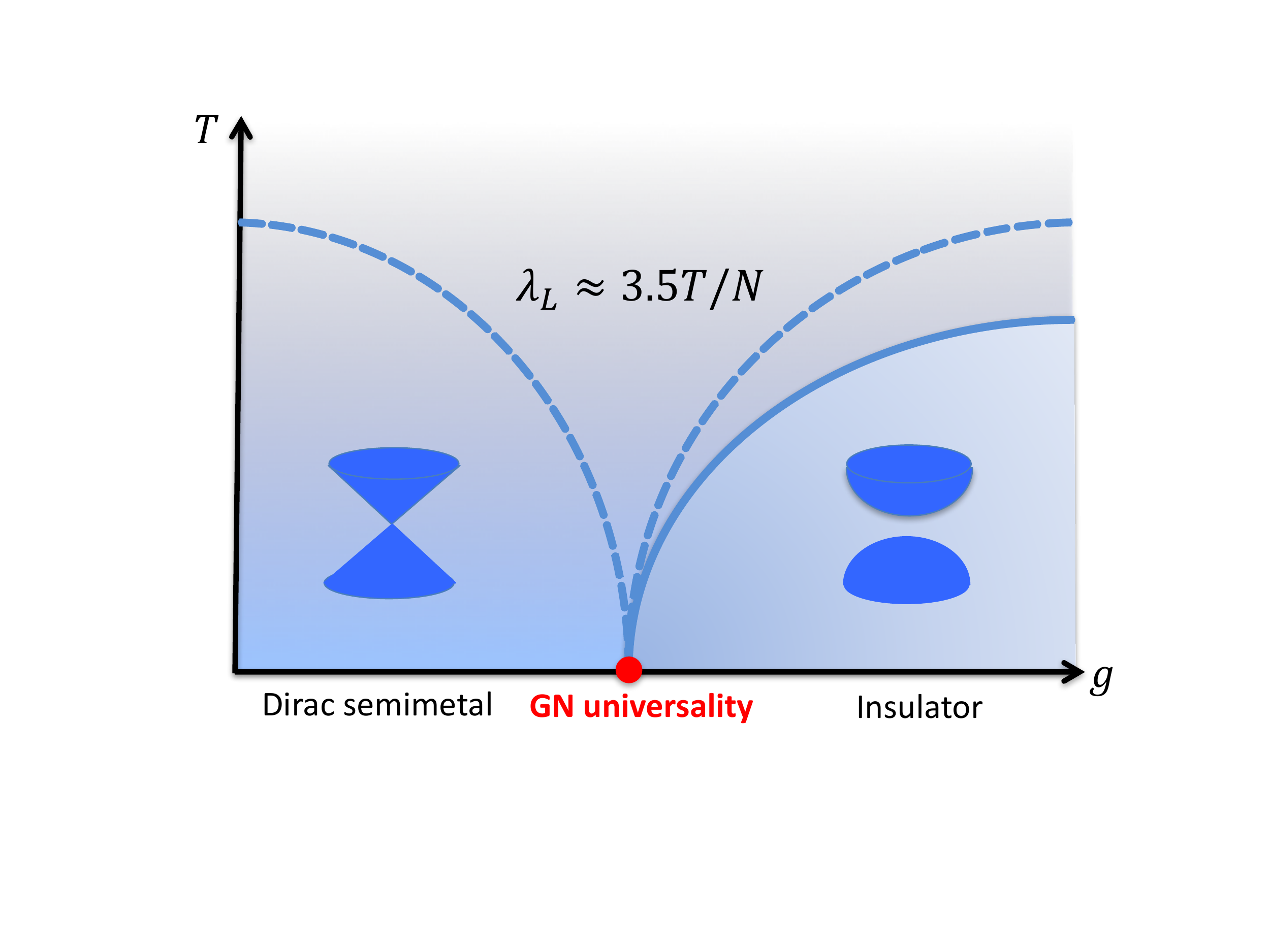}
\caption{The schematic phase diagram of the GN models in (2+1) dimensions, where the Lyapunov exponent in the quantum critical regime is indicated. The dashed an solid lines indicate quantum critical regime and finite temperature transition, respectively.}\label{phasediagram}
\end{figure}

\section{Preliminaries}
The (2+1)-dimensional GN model is defined by the (Euclidean) Lagrangian:
\bea
	\mathcal{L} = \sum_i \psi_i^\dag (\partial_\tau - i \vec \sigma \cdot \nabla) \psi_i - \frac{g}{4N} (\sum_i \psi_i^\dag \sigma^z \psi_i)^2,
\eea
where the summation range is $i=1,...,N$, and each $\psi_i$ refers to a two-component Dirac spinor, $g$ is the interaction strength. In the following, the summation over flavor will be implicit. $\tau$ denotes the imaginary time, $\vec \sigma \equiv (\sigma^x, \sigma^y)$, and $\nabla \equiv (\partial_x, \partial_y)$. We have set the Fermi velocity to one for simplicity. The model is invariant under various reflection transformations, e.g., $x \rightarrow -x$, $\psi_i \rightarrow i \sigma_x \psi_i$.

After introducing a real boson $\phi$ to decouple the four-fermion interaction, we obtain
\bea\label{GN}
	\mathcal{L} = \psi^\dag (\partial_\tau - i \vec \sigma \cdot \nabla) \psi + \frac1g \phi^2 + \frac1{\sqrt{N}} \phi \psi^\dag \sigma^z \psi.
\eea
Under the symmetry transformation defined as $\phi \rightarrow -\phi$, $x\to -x$, and $\psi_i\rightarrow i\sigma_x\psi_i$, the Lagrangian in \Eq{GN} is invariant. Thus a nonzero condensate of the boson field signals a spontaneous $Z_2$ symmetry breaking. At large-$N$ limit, the gap equation is given by $\frac1g=\Omega(m)$, where $m$ is the order parameter, (i.e., $m= \frac{\langle \phi \rangle}{\sqrt{N}}$), and $\Omega(m)=\int_k \frac1{\omega_n^2+ k^2 + m^2} $, where $\omega_n = (2n+1)\pi T  $ is the fermionic Matsubara frequency, $k=|\vec k|$, and $\int_k \equiv T \sum_n \int_{\vec k}$, $\int_{\vec k} \equiv \int \frac{d^2k}{(2\pi)^2}$ (see the Appendix for details). Then $\frac1g=\Omega(0)$ defines the transition points, i.e.,
\bea
	\frac1{g_c}- \frac1{g_c(T)}= \frac{\log 2}{2\pi} T,
\eea
where $g_c(T)$ refers to critical interaction strength in the limit of large-$N$ at temperature $T$ and $g_c$ is the coupling strength at quantum critical point, i.e., $g_c \equiv g_c(0)$. See the Appendix for details. At temperature $T$, when $g>g_c(T)$, the Ising symmetry is spontaneous broken. In the following, we set $g=g_c$ to explore the properties of quantum critical theory.

\begin{figure}[t]
\includegraphics[width=7cm]{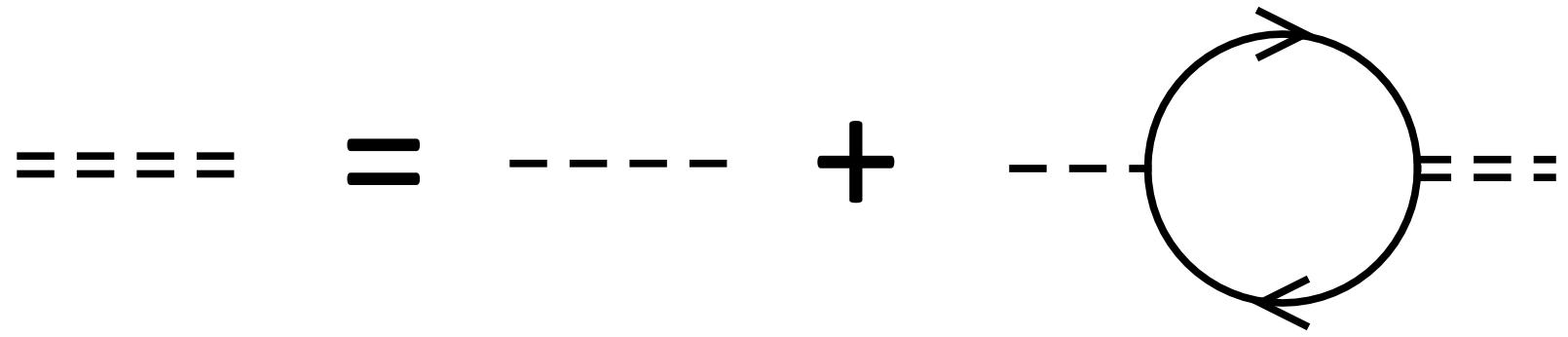}
\caption{\label{polarization} Feynman diagram representation of the Dyson-Schwinger equation for the boson propagator. The diagram shows the lowest nontrivial order in large-$N$ expansion. The single (double) dashed line represents bare (dressed) propagator of bosons, while the solid arrowed line represents the bare propagator of fermions. The polarization function is of order one, because summation over $N$ flavors of fermions cancels $(\frac1{\sqrt{N}})^2$ coming from the two vertices.}
\end{figure}

The imaginary time ordered propagators are defined as $G(\tau, \vec x)= \langle \mathcal{T}_\tau \psi(\tau, \vec x) \psi^\dag \rangle$, and $D(\tau, \vec x)= \langle \mathcal{T}_\tau \phi(\tau, \vec x) \phi \rangle$, where $\psi$ and $\phi$ imply $\psi(0,\vec 0)$ and $\phi(0, \vec 0)$, respectively. The retarded Green's functions are defined as $G_R(t, \vec x)= -i\theta(t) \langle \{\mathcal{T}_\tau \psi(\tau, \vec x), \psi^\dag \} \rangle$, and $D_R(\tau, \vec x)= -i \theta(t) \langle [\phi(\tau, \vec x) ,\phi] \rangle$, where $\theta$ denotes the step function, and $[,]$ and $\{,\}$ refer to commutator and anti-commutator, respectively. These Green's functions are related by the analytical continuation, $G_R(\omega, \vec k) = -G(\omega+ i\delta, \vec k)$, where $G(i \omega_n, \vec k) \equiv \int_k G(\tau, \vec k) e^{-i \omega_n \tau + i \vec k \cdot \vec x}$. The bare propagators of fermions and bosons are $G_0= (-i\omega_n+ \vec k\cdot \vec \sigma)^{-1}$ and $D_0= g/2$. The spectrum function can be obtained from the retarded Green's function through $A(\omega, \vec k)= -2 \text{Im} G_R(\omega, \vec k)$ for fermions while $A_\phi(\omega,\vec k)=-2\text{Im} D_R(\omega,\vec k)$ for bosons.

Since we are interested in calculating the OTOC which involves the Green's function defined in both real and imaginary times, we introduce the Wightman propagators $G_W(t, \vec x)= Z^{-1} \text{Tr}[\sqrt{\rho} \phi(t, \vec x) \sqrt{\rho} \phi]$, where $\rho= e^{-H/T}/Z$. The Wightman propagators of  fermions and bosons are related to their spectrum functions via:
\bea
	G_W( \omega, \vec k) = \frac{A(\omega, \vec k)}{2\cosh \frac{\omega}{2T}},\quad D_W( \omega, \vec k)= \frac{A_\phi(\omega, \vec k)}{2\sinh \frac{\omega}{2T}} \\
	\nn
\eea
where $A$ and $A_\phi$ refer to the spectrum functions of fermions and bosons, respectively.

\section{Polarization function at finite temperature}
Owing to the short-range interactions of fermions, the bare boson propagator is just a constant, $D_0(i \omega_n, \vec k) \equiv \frac{g}2$. As shown in Fig. \ref{polarization}, according to Dyson-Schwinger equation, the dressed propagator is given by
\bea
	D(i\omega_n, \vec k)= \frac{1}{\frac2g- \Pi(i\omega_n, \vec k)},
\eea
where $\Pi$ refers to the polarization function. To the lowest order at the large-$N$ limit, the dynamic polarization function is
\bea
	\Pi(i\nu_n, \vec p)= - \int_k \text{Tr} [G_0(i\omega_n-i\nu_n , \vec k-\vec p) \sigma^z G_0(i\omega_n , \vec k) \sigma^z], \nn \\
\eea

At zero-temperature, the polarization can be calculated via Feynman trick, $ \Pi(i\nu_n, \vec p)= -\frac1{8} \sqrt{\nu_n^2+p^2}$, where $p \equiv |\vec p|$. After analytic continuation, the dressed retarded polarization function at zero temperature takes the form:
\bea\label{polarization_eqn1}
	\Pi_R(\nu, \vec p) &=& -\frac1{8} \sqrt{p^2-\nu^2} \theta(p-|\nu|) \nn \\
		&& + \frac{i \text{sgn}(\nu)}{8}  \sqrt{\nu^2- p^2} \theta(|\nu|-p).
\eea

\begin{figure}[t]
	\subfigure[$\ I_1(\Omega, P)$]{
		\includegraphics[height=3.1cm]{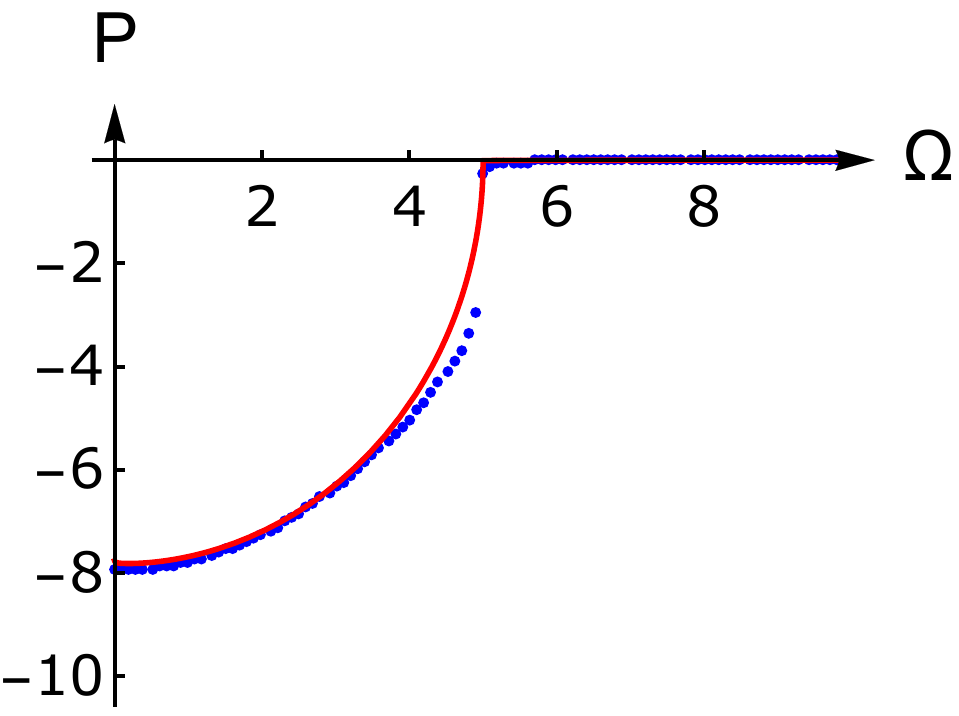}}
	\quad
	\subfigure[$\ I_2(\Omega, P)$]{
		\includegraphics[height=3.3cm]{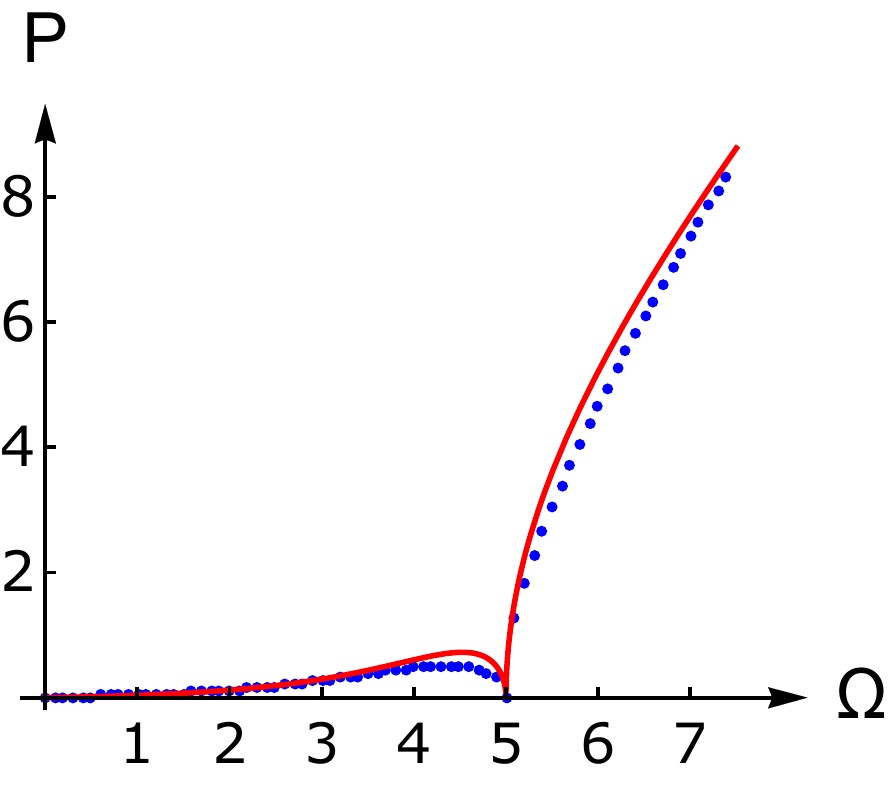}}
	\caption{\label{comparison} Comparisons between the numerical evaluations of the dimensionless functions $I_i(\Omega, P)$ and the analytic approximations in Eq. (\ref{polarization_eqn2}).}
\end{figure}

However, there is no simple formula at finite temperature. To simplify the calculation, we use the projection operators that corresponding to the helical basis of the Hamiltonian $  (\vec k \cdot \vec \sigma) \mathcal{P}_a(\vec k)= a k \mathcal{P}_a(\vec k)$, where $\mathcal{P}_a(\vec k)= \frac12 (1+ a \hat k \cdot \vec \sigma)$, and $\hat k \equiv \vec k/ k$, $a=\pm 1$. Plugging the fermion propagators $G_0(i\omega_n , \vec k)= \sum_a \mathcal{P}_a(\vec k)/(- i\omega_n + a k)$ into the polarization, and continuing to real frequency, i.e., $\Pi_R(\nu, \vec p)= \Pi(\nu+ i\delta, \vec p)$, we get
\bea
	&& \Pi_R(\nu, \vec p)= \sum_{a,b} \int_{\vec k} K_{ab}(\vec k, \vec k- \vec p) \frac{n(b|\vec k|)-n(a|\vec k-\vec p|)}{\nu+ a|\vec k-\vec p|- b|\vec k|+ i\delta } \nn \\
		&& = \sum_{a,b} \int_p^\infty \frac{d\xi}{2\pi} \int_{0}^{p} \frac{d\eta}{2\pi} \Big( \frac{p^2-\eta^2}{\xi^2-p^2} \Big)^{ab/2} \frac{n(b \frac{\xi+\eta}2)-n(a \frac{\xi-\eta}2)}{\nu+ a\frac{\xi-\eta}2- b\frac{\xi+\eta}2+ i\delta }, \nn \\
\eea
where $n(x)=1/[\exp(\frac{x}T)-1]$ is the Bose-Einstein distribution resulted from Matsubara frequency summation. In the calculations, we have used
\bea
	\text{Tr}[\mathcal{P}_a(\vec k) \sigma^z \mathcal{P}_b(\vec p) \sigma^z] = K_{ab}(\vec k, \vec p), ~~ K_{ab}(\vec k, \vec p)= \frac{1- ab \hat k\cdot \hat p}2, \nn \\
\eea
with $\hat k \equiv \vec k/k$, and made the variable transformations $\hat k\cdot \hat p= \frac{k^2+p^2-q^2 }{2kp}$ and $\xi= k+ q, \eta= k- q$ (see the Appendix for details). These integrals can be simplified further to two dimensionless functions,
\bea
\Pi_R'(\nu, \vec p)= \frac{T}{2\pi} I_1 \Big( \Omega, P \Big), \quad \Pi_R''(\nu, \vec p)= \frac{T}{2\pi} I_2 \Big( \Omega, P \Big), \nn \\
\eea
where $\Omega= \frac{\nu}{2T}$ and $P= \frac{p}{2T}$ are dimensionless argument, $\Pi_R'$ ($\Pi_R''$) is the real (imaginary) part of $\Pi_R$, and $I_i$ are defined in the Appendix. At critical point, we have absorbed the critical interaction strength to real part of the polarization function, i.e., $\Pi_R' \rightarrow \Pi_R'- \frac2{g_c}$, thus $ D_R(\nu, \vec p) = \Pi_R(\nu, \vec p)^{-1} $.

For $P\equiv \frac{p}{2T} \gg 1$, the approximated polarization function at finite temperature is given by 
\begin{widetext}
\bea\label{polarization_eqn2}
\Pi_R(\nu, p)= \frac{p}{4\pi } \!\times\! \begin{cases} \frac{2}\pi K_0(P) h_1(\nu/p)- \frac\pi2 \sqrt{1-(\nu/p)^2} + i  e^{P(\nu/p)} K_0(P) \sqrt{1-(\nu/p)^2}, \quad &0<\nu<p \\
 \\
	\frac{2}\pi K_0(P) h_2(\nu/p)+ i \frac\pi2 \sqrt{(\nu/p)^2-1}, & 0<p<\nu
\end{cases}
\eea
\end{widetext}
where $K_0$ is the Bessel function of the second type which has the property $K_0(x) \approx \sqrt{\frac\pi{2x}} e^{-x}$ for $x \gg 1$, and $h_1, h_2$ are two dimensionless functions (see the Appendix). Thus, in $\nu> p$ ($\nu<p$) region, the real (imaginary) part of polarization function is exponentially suppressed. At zero-temperature limit, $P \rightarrow \infty$, Eq. (\ref{polarization_eqn2}) recovers Eq. (\ref{polarization_eqn1}). To justify our approximations, we also plot the comparisons between the numerical evaluations of the dimensionless functions $I_i(\Omega, P)$ and the analytic approximations, as shown in Fig. \ref{comparison}.

We further define two dimensionless propagators that will be useful later:
\bea
	D_W(\Omega, P) &=& \frac{1}{\sinh \Omega} \frac{I_2(\Omega, P)}{I_1(\Omega, P)^2+ I_2(\Omega, P)^2}, \\
	|D_R(\Omega, P)|^2 &=& \frac{1}{ I_1(\Omega, P)^2+ I_2(\Omega, P)^2}.
\eea
Note that we will use capital arguments to indicate these dimensionless propagators.

\section{Quantum scattering rate}
\begin{figure}[t]
	\includegraphics[width=6cm]{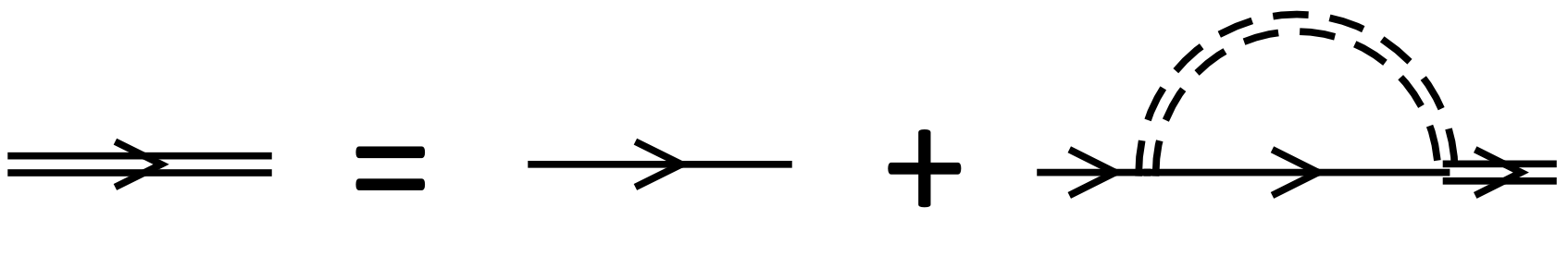}
	\caption{\label{selfenergy}Feynman diagram representation of the Dyson-Schwinger equation for the fermion propagator. The diagram shows the lowest nontrivial order in large-$N$ expansion. The single (double) arrowed line represents bare (dressed) propagator of fermions. The polarization function is of order $1/N$ coming from  the two vertices.}
\end{figure}

Now we consider the self-energy of Dirac fermions resulted from coupling to the dynamic bosons, which is given by
\bea
	\Sigma(i \nu_n, \vec p)= \frac1{N} \int_k  \sigma^z G_0(i \omega_m - i \nu_n, \vec k -\vec p) \sigma^z D(i \omega_m, \vec k). \nn \\
\eea
where $\omega_m =2m \pi T $ and $\nu_n=(2n+1) \pi T$ are Matsubara frequencies. As indicated in Fig. \ref{selfenergy}, we use bare fermion propagator and dressed boson propagator in calculating the fermionic self-energy.

As shown in Fig. \ref{selfenergy}, the Dyson-Schwinger equation reads $G^{-1}= G_0^{-1} + \Sigma$. In the helical basis, the retarded Green's function takes the form $G_{R,a}^{-1}(\nu, p)\approx \nu- \nu^\ast+ i \Gamma_{a,p}$, where $\nu^\ast= a p- \Sigma'_{R,a}(ap, p)$ is the renormalized energy, and $ \Gamma_{a, p} = \Sigma''_{R,a}(ap,p)$ is the quantum scattering rate. $\Sigma_{a,R}(\omega, \vec k)= \Sigma_a(\omega+ i\delta, k) $ is the retarded self-energy in the helical basis. And $\Sigma'_{R,a}$ ($ \Sigma''_{R,a}$) is the real (imaginary) part of retarded self-energy. Projecting the self-energy into the helical basis, we have
\bea
	&&\Sigma_a(i \nu_n, \vec p) = \frac1{N} \sum_b \int_k  \frac{K_{ab}(\vec p, \vec k- \vec p)D(i \omega_n, \vec k)}{-i (\omega_n -\nu_n)+b| \vec k -\vec p|}   \nn \\
	&&= -\frac1{N} \sum_b \int_{\vec k} K_{ab}(\vec p, \vec k- \vec p) \Big[n_F(b |\vec k - \vec p|) D(b|\vec k - \vec p|+ i\nu_n, \vec k) \nn \\
	&& ~~~~~~~~~~~~~~~~~~ +  \int \frac{dx}{2\pi i} \frac{n(x)[D(x+ i\delta, \vec k)-D(x- i\delta, \vec k)]}{x- i\nu_n - b |\vec k - \vec p| }  \Big]  \nn\\
	&&= -\frac1{N} \sum_b \int_{\vec k} K_{ab}(\vec p, \vec k- \vec p) \Big[n_F(b |\vec k - \vec p|) D(b|\vec k - \vec p|+ i\nu_n, \vec k) \nn \\
	&& ~~~~~~~~~~~~~~~~~~ -  \int \frac{dx}{2\pi } \frac{n(x) A_\phi(x, \vec k )}{x- i\nu_n - b |\vec k - \vec p| }  \Big],
\eea
and the imaginary part is given by:
\begin{widetext}
\bea
 \Sigma''_{R,a}(\nu, \vec p) &=&\frac1{N} \sum_b \frac12  \int_{\vec k}  K_{ab}(\vec p, \vec k- \vec p)[ n(b |\vec k - \vec p|+ \nu)+ n_F(b |\vec k - \vec p|)] A_\phi(\nu+ b|\vec k - \vec p|, \vec k) \nn \\
	&=& \frac1N \sum_b \frac1{2p}\int_0^\infty \frac{kdk}{2\pi} \int_{|p-k|}^{p+k} \frac{dq}{2\pi} \Big( \frac{k^2-(p-q)^2}{(p+q)^2-k^2} \Big)^{ab/2} A_\phi(bq+\nu, k) [n_F(bq)+ n(bq+\nu)] \nn \\
	&=& \frac1N\sum_b \frac{\cosh \frac{\nu}{2T}}{2p} \int_0^\infty \frac{kdk}{2\pi} \int_{|p-k|}^{p+k} \frac{dq}{2\pi} \Big( \frac{k^2-(p-q)^2}{(p+q)^2-k^2} \Big)^{ab/2} \frac{D_W(bq+\nu, k)}{\cosh \frac{q}{2T}},
\eea
\end{widetext}
where $n_F(x) \equiv 1/[\exp(\frac{x}T)+1]$ is the Fermi-Dirac distribution. By inverse-Fourier transforming to time domain, it is straightforward to see that the renormalized energy, $\nu^\ast$, corresponds to the frequency of wave function; while the quantum scattering rate, $\Gamma_p$, is the inverse lifetime. Since it encodes the dynamic properties of the system, we focus on the quantum scattering rate. According to particle-hole symmetry, $\Gamma_{p,+}=\Gamma_{p,-}$, we calculate $\Gamma_{p}= \sum_a \Gamma_{p,a} =\frac{4 T}N \Gamma_P$, where $\Gamma_P$ is a dimensionless function of $P\equiv \frac{p}{2T}$:
\bea\label{scatteringrate_eqn}
	&& \Gamma_{P} =  \sum_{b=\pm} \int_0^\infty KdK \int^{K+P}_{|K-P|} \frac{dQ}{2\pi} \nn \\
	&& ~~~\times \frac1P \frac{\cosh P}{\cosh Q} \sqrt{\frac{K^2-(P-bQ)^2}{(P+bQ)^2-K^2}} D_W(P+b Q, K).~~~
\eea

\begin{figure}
	\includegraphics[width=5cm]{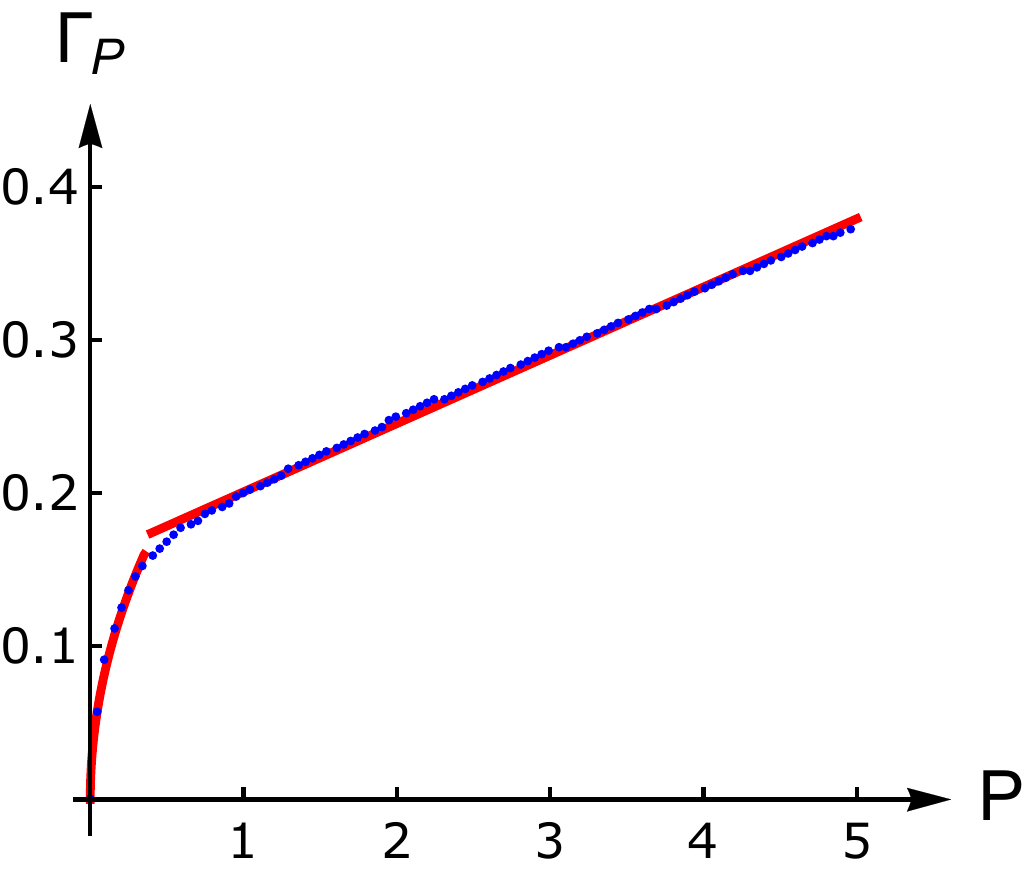}
	\caption{\label{scatteringrate}The dimensionless function $\Gamma_P$. The blue dots represent the numerical values of $\Gamma_P$. The red curves represents the fitted ones. We set the cutoff of integration by $\Lambda=8$ in the numerical calculation, which means we conclude the energies up to $8T$. For $P\ll 1$, it is proportional to $\sqrt{P}$ as indicated by the fitted curve given by $\sqrt{P}$. For $P \sim 1$, we can see that $\Gamma_P$ is almost linear in $P$.}
\end{figure}

In the low energy region, $P\ll1$, where the momentum (energy) is much smaller compared to the temperature, we have $\Gamma_P \propto \sqrt{P}$ (see the Appendix for detailed calculations). As a result the quantum scattering rate reads $\Gamma_{p} \propto  \frac{T}N \sqrt{P}=\frac{T}N \sqrt{\frac{p}{2T}}$ . Thus at low-energy limit, the excitation with energy $p$ will have the profile like $e^{ip t- \frac{C\sqrt{Tp}}{N} t }$ in the time domain, with $C$ a positive constant. If $p< \frac{C^2}{N^2} T$, the excitation will be destroyed before it propagates a full wavelength; that means the quasiparticle picture breaks down. On the other hand, when the energy is comparable to the temperature, we are not able to get an analytic result from Eq. (\ref{scatteringrate_eqn}). 
Instead, by numerically calculating the dimensionless function $\Gamma_P$, we find that it is linear in $P$
as the excitation energy is comparable to the temperature, as shown in Fig. \ref{scatteringrate}, indicating the quantum scattering is proportional to the energy of the excitations. In Fig. \ref{scatteringrate}, we also verify that the quantum scattering rate is proportional to $\sqrt{p}$ in the low energy limit.

\begin{figure}[t]
	\includegraphics[width=9.cm]{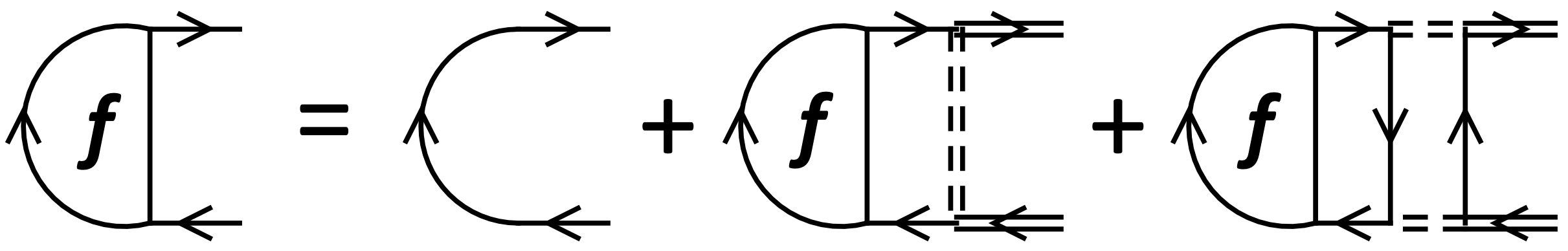}
	\caption{\label{OTOC} Feynman diagram representation of the Bether-Saltpeter equation for the out-of-time-order correlator. The diagram shows the lowest nontrivial order in large-$N$ expansion. All diagrams are of order $1/N$. Note that. In the real-time contour, we use the dressed propagators for both bosons and fermions. In the imaginary-time contour, we use dressed boson Wightman propagator and bare fermion Wightman propagator, because the dressed fermion propagator only leads to higher order corrections in $1/N$.}
\end{figure}

\section{Out-of-time-order correlator}
As the basic degrees of freedom in the GN model are the Dirac fermions, the quantity that captures information scrambling is the (squared) norm of anti-commutators:
\bea
	C(t)=\frac1{N^2} \sum_{ij,\alpha}\int d^2 x \text{Tr}(|\{\psi_{i\alpha}(t, \vec x), \psi_{j}^{\dag\alpha}\}|^2 \rho).
\eea
where $\rho= e^{-H/T}/Z$. The factor $1/N^2$ is to properly normalize the summation over $i,j$. To get the Bether-Saltpeter equation, we keep two spin indices fixed as $\alpha$ and $\beta$:
\bea
	&& f_\alpha^{\ \beta}(t) = \nn \\
	&& \frac1{N^2} \sum_{ij,\gamma}\int d^2 x \text{tr}(\{\psi_{i\alpha}(t, \vec x), \psi_{j}^{\dag\gamma}\} \sqrt{\rho} \{\psi_{j\gamma},\psi^{\dag\beta}_{i}(t, \vec x)\} \sqrt{\rho}). \nn \\
\eea

Since we will calculate the Feynman diagrams in frequency space, it is illuminating to see how Lyapunov exponent manifests in frequency domain. 
The exponential growth of OTOC indicates the differential equation, $\partial_t C(t) = \lambda_L C(t)$. Fourier transforming to frequency domain leads to $-i\nu C(\nu) = \lambda_L C(\nu)$, where $C(\nu)$ is the OTOC in frequency domain. 
As shown in Fig. \ref{OTOC}, $f_\alpha^{\ \beta}(\nu)$ can be obtained by $f_\alpha^{\ \beta}(\nu) = \int_k f_\alpha^{\ \beta}(\nu; \omega, \vec k) $. And $f_\alpha^{\ \beta}(\nu; \omega, \vec k)$ satisfies the Bether-Saltpeter equation:
\bea\label{OTOC_eqn}
	&& f_\alpha^{\ \beta}(\nu; \omega, \vec k) =   \frac1N G_R(\omega+\nu, \vec k)_\alpha^{\ \gamma}G_A(\omega, \vec k)_\delta^{\ \beta} \nn \\	
	&& ~~~~~~\times \Big[ \delta_\gamma^\delta + \int_{k'}  \Gamma^{\delta\gamma'}_{\gamma\delta'}(\nu; \omega, \vec k,\omega', \vec k')f_{\gamma'}^{\ \delta'}(\nu; \omega', \vec k') \Big],
\eea
where the summation over spin index is implicit and $\Gamma^{\delta\gamma'}_{\gamma\delta'}$ indicates the kernel function. As shown in Fig. \ref{OTOC}, to the order of $1/N$, the kernel consisting of two parts:
\begin{widetext}
\bea
	&& \Gamma^{\delta\gamma'}_{\gamma\delta'}(\nu; \omega, \vec k,\omega', \vec k') = (\sigma^z)_\gamma^{\ \gamma'} (\sigma^z)_{\delta'}^{\ \delta} D_W(\omega-\omega', \vec k- \vec k') \nn \\
	&& ~~~~~~~~~~~~~ +\int_{k''} [\sigma^z G_W(\omega- \omega'', \vec k- \vec k'') \sigma^z]_\gamma^{\ \delta} [\sigma^z G_W(\omega'- \omega'', \vec k'- \vec k'') \sigma^z]^{\ \gamma'}_{\delta'} D_R(\omega''+\nu, \vec k'') D_A(\omega'', \vec k''),
\eea
\end{widetext}
where the first and second term correspond to one-rung kernel and two-rung kernel, respectively.

The product of retarded and advanced propagators can be simplified as
\bea
	 G_R(\omega+\nu, \vec k)_\alpha^{\ \delta}G_A(\omega, \vec k)_\gamma^{\ \beta} =\sum_{a,b}  \frac{\mathcal{P}_a(\vec k)_\alpha^{\ \delta} \mathcal{P}_b(\vec k)_\gamma^{\ \beta}}{(\omega+\nu-a k)(\omega-b k)}  \nn \\
	 \approx \sum_{a} \mathcal{P}_a(\vec k)_\alpha^{\ \delta} \mathcal{P}_a(\vec k)_\gamma^{\ \beta}  \frac{2\pi i\delta(\omega-a k)}{\nu+i\delta}. \nn \\
\eea
Here, we have neglected the contribution away from zero frequency, such as $1/(\nu \pm 2k + i\delta)$, because we are interested in the increasing part rather than the oscillating part of OTOC \cite{swingle2017a}. After taking the quantum scattering rate into consideration, we have
\bea
	&&G_R(\omega+\nu, \vec k)_\alpha^{\ \delta}G_A(\omega, \vec k)_\gamma^{\ \beta} \nn \\
	&& \quad\quad\quad= \sum_{a} \mathcal{P}_a(\vec k)_\alpha^{\ \delta} \mathcal{P}_a(\vec k)_\gamma^{\ \beta}  \frac{2\pi i\delta(\omega-a k)}{\nu+i2 \Gamma_{k, a}}.
\eea
The zeroth-order OTOC is then $f_\alpha^{\ \beta}(\nu; \omega, \vec k)=\sum_{a} \mathcal{P}_a(\vec k)_\alpha^{\ \beta} \frac{2\pi i \delta(\omega-a k)}{\nu+i2\Gamma_{k, a}}$. Thus we use the following ansatz for OTOC,
\bea
	f_\alpha^{\ \beta}(\nu; \omega,\vec k)=\sum_{a} f_a(\nu,\vec k) P_a(\vec k)_\alpha^{\ \beta} 2\pi \delta(\omega-a k).
\eea
where $f_a(\nu;\vec k)$ is the component in helical basis. The ansatz fixes the dependence of frequency $\omega$, and leads to
\bea
	&&(-i \nu+ 2\Gamma_{a,k}) f_a(\nu; \vec k)\nn \\
	&& ~~~~~~~~~= \frac1N \sum_b \int_{\vec k'}[ M_{ab}^{(1)}(\vec k, \vec k')+ M_{ab}^{(2)}(\vec k, \vec k') ] f_b(\nu; \vec k'), \nn \\
\eea
where we have neglected the first term in Eq. (\ref{OTOC_eqn}), because it is not important to the exponential increasing part \cite{swingle2017a, schmalian2017}. Here, $M_{ab}^{(1)}$ and $M_{ab}^{(2)}$ are the one-rung and the two-rung kernels, respectively:
\bea
	&& M_{ab}^{(1)}(\vec k, \vec k')=K_{ab}(\vec k, \vec k') D_{W}(a k-b k', \vec k- \vec k'), \\
	&& M_{ab}^{(2)}(\vec k, \vec k')=\! \sum_{a'b'} \!\int_{k''} K_{aa'}(\vec k, \vec k\!-\! \vec k'') G_{W,a'}(a k \!-\!\omega'', \vec k \!-\! \vec k'') \nn\\
	&& ~~\times K_{bb'}(\vec k', \vec k' \!-\! \vec k'')  G_{W,b'}(b k'- \omega'', \vec k' \!-\! \vec k'') |D_R(\omega'', \vec k'')|^2. \nn \\
\eea
Note in $M_{ab}^{(2)}$, we have approximated $D_R(\omega''+\nu, \vec k'') D_A(\omega'', \vec k'') \approx |D_R(\omega'', \vec k'')|^2$, because it only affects higher orders in $1/N$. In the calculations, we will use the bare Wightman propagator for fermions, i.e., $G_{W,a}(\omega, \vec k) \approx \frac{\pi \delta(\omega- a|\vec k|)}{\cosh \frac{\omega}{2T} }$. More specifically, in the real-time contour, we use the dressed propagators for both bosons and fermions. While in the imaginary-time contour, we use dressed boson Wightman propagator and bare fermion Wightman propagator, because the dressed fermion propagator only leads to higher order corrections in $1/N$.

Now assuming the OTOC is rotationally invariant, (actually we only need the assumption that the eigenfunction corresponding to the largest eigenvalue of kernel function is rotationally invariant), we can integrate over angle first:
\bea
	&& (-i \nu+ 2 \Gamma_{k, a}) f_a(\nu; k) \nn \\
	&&~~~~~ = \frac1N\sum_b \int \frac{k'dk'}{2\pi} [\mathcal{M}_{ab}^{(1)}( k, k')+\mathcal{M}_{ab}^{(2)}(k, k')] f_b(\nu; k'). \nn \\
\eea
where $\mathcal{M}_{ab}^{(i)}(k, k')$ are the resulted kernel functions after angle integration.

Owing to the particle-hole symmetry, $\mathcal{M}_{ab}$ reduce to two independent parts.
\bea
\mathcal{M}_{++}^{(i)}(k, k')= \mathcal{M}_{--}^{(i)}(k, k'),  \quad \mathcal{M}_{+-}^{(i)}(k, k')= \mathcal{M}_{-+}^{(i)}(k, k'). \nn \\
\eea
So we define $\mathcal{M}_{+}^{(i)}(k, k')\equiv \mathcal{M}_{++}^{(i)}(k, k')$, and $ \mathcal{M}_{-}^{(i)}(k, k')\equiv \mathcal{M}_{+-}^{(i)}(k, k')$, which indicate band-preserving and band-changing parts. The explicit form of the kernel functions, which are very complicated, are given in the Appendix. Thanks to the particle-hole symmetry, we can consider the sum of the two components of OTOC in helical basis, $f(\nu; k) = \sum_a f_a (\nu; k)$:
\bea\label{kernel1}
(-i \nu+ 2 \Gamma_{k}) f(\nu; k) = \frac1N \sum_{a,i} \int \frac{k'dk'}{2\pi} \mathcal{M}_{a}^{(i)}( k, k')f(\nu; k'). \nn \\
\eea
However, because of the complicated form of the kernel functions, it is not possible to diagonalize them analytically. We will numerically diagonalize the kernel functions and get the Lyapunov exponent in the next section.

\section{Many-body quantum chaos}
By making the quantum scattering rate in Eq. (\ref{kernel1}) a part of the kernel functions, one can see that $\nu$ is given by the eigenvalues of the kernels:
\bea\label{f}
	&& -i \nu f(\nu; k)  = \nn \\
	&& \frac1N \int \frac{k'dk'}{2\pi} [\sum_{a,i} \mathcal{M}_{a}^{(i)}( k, k') - 4\pi \delta(k-k') \Gamma_{k}] f(\nu; k'). \nn\\
\eea
Especially, the largest eigenvalue gives rise to the maximal Lyapunov exponent. However, the kernel functions in Eq. (\ref{f}) is not symmetric in their arguments. By multiplying $\sqrt{k k'}$ in both hand side, we make them symmetric for the convenience of numerical diagonalization. To facilitate the numerical calculation, we further make the kernel functions dimensionless, i.e.,
\bea
	&& -i \nu \sqrt{K} f(\nu; K) = \nn \\
	&& \frac{4T}N  \int dK'  \Big( \sum_{i,a} \mathcal{F}_{a}^{(i)}( K, K')- \delta(K-K') \Xi_K \Big)\sqrt{K'} f(\nu; K'), \nn \\
\eea
where one-ladder kernels are
\bea
&& \mathcal{F}_{\pm}^{(1)}(K, K') = \nn \\
 &&\frac1{\sqrt{KK'}} \int_{|K-K'|}^{K+K'} \frac{QdQ}{2\pi} \sqrt{ \frac{Q^2-(K\mp K')^2}{(K\pm K')^2-Q^2}}D_{W}(K\mp K', Q) , \nn \\
\eea
and two-rung kernels are
\bea
&&\mathcal{F}_{\pm}^{(2)}(K, K') = \frac{4\pi}{\sqrt{KK'}}\int \frac{QdQ}{2\pi}  \int \frac{d\Omega}{2\pi} \sqrt{\frac{Q^2-\Omega^{2}}{(2K-\Omega)^2-Q^2}} \nn \\
&& ~~~\times \sqrt{\frac{Q^2-\Omega^{2}}{(2K' \mp \Omega)^2-Q^2}}  \frac{|D_{R}(\Omega, Q)|^2}{\cosh (K-\Omega) \cosh (K' \mp \Omega)}.
\eea
Note that the integration ranges of $\Omega$ in $\mathcal{F}^{(2)}_{\pm}$ are different: for $\mathcal{F}^{(2)}_{+}$, the range is $(-Q,\text{min}[K\!-\!|Q\!-\!K|,K'\!-\!|Q\!-\!K'|]) \cup (\text{max}[K\!+\!|Q\!-\!K|,K'\!+\!|Q\!-\!K'|],\text{min}[Q\!+\!2K,Q\!+\!2K'])$; while for $\mathcal{F}^{(2)}_{-}$, it is $(-Q,\text{min}[K\!-\!|Q\!-\!K|, \!-\!K'\!-\!|Q\!-\!K'|])\cup (\text{max}[K\!+\!|Q\!-\!K|, |Q\!-\!K'|\!-\!K'],Q)$. Finally, the part corresponding to the quantum scattering rate is given by
\bea
	&&\Xi_K=4\pi \sum_a \int \frac{PdP}{2\pi} \int_{|P-K|}^{P+K} \frac{dQ}{2\pi} \nn \\
	&& ~~~\times \frac{\cosh K }{\cosh Q} \sqrt{\frac{P^2-(K-aQ)^2}{(K+aQ)^2-P^2}} D_W(K+aQ, P),
\eea
where $D_{W}(\Omega, Q)$ and $|D_{R}(\Omega, Q)|^2$ are the dimensionless propagators introduced in Sec. II. Now the kernel functions are symmetric under exchange of $K$ and $K'$ and the eigenfunction changes to the form $\sqrt{K} f(K)$, but the eigenvalues remain unchanged.

\begin{figure}
	\includegraphics[width=6cm]{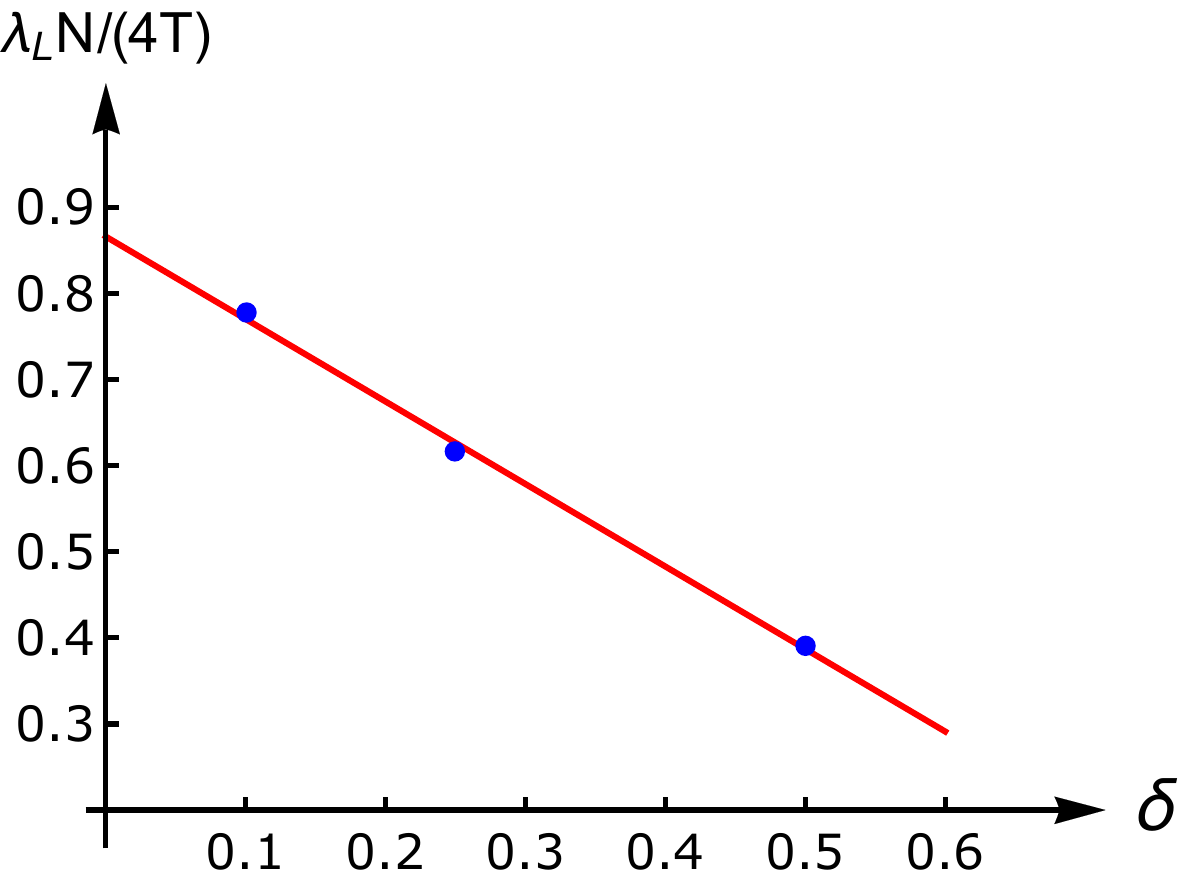}
	\caption{\label{extrapolation} The extrapolation of the Lyapunov exponent as a function of the discretized interval characterized by $\delta$. The interception is $0.87$, giving rise to the eigenvalue extrapolating to continuous function. In this plot, the cutoff of the integral in the calculations is set by $\Lambda=12$. Up to this scale $\Lambda=12$, the dependence of Lyapunov exponent on cutoff is negligible.}
\end{figure}

In the numerical calculation, we discretize the argument $K$ and $K'$ in kernel functions and replace the integral by summation over small intervals with the length $\delta$. The discretization is characterized by $\delta$. After the discretization and evaluation of the integrals, the kernel functions become a symmetric matrix of $K, K'$ that can be diagonalized directly, and the Lyapunov exponent is the largest eigenvalue. We extrapolate the Lyapunov exponent of each discretization $\delta$, as shown in Fig. \ref{extrapolation}, to get the result $\lambda_L \approx 0.87\times \frac{4T}{N} \approx 3.5 \frac{T}{N}$. The cutoff of the integral in the calculations of Fig. \ref{extrapolation} is set to be $\Lambda=12$, which means the cutoff of energy is up to 24 times the temperature. We have also increased the cutoff to $\Lambda=24$ and $\Lambda=36$, and find the same answer up to two digits, i.e., the dependence of the Lyapunov exponent on cutoff is negligible at this scale.

\section{Discussion}
In this paper, we have computed the quantum scattering rate and the Lyapunov exponent at the quantum criticality of the GN models to the leading order of large-$N$ expansion. The GN model can characterize the critical phenomena of the $Z_2$ transitions in a (2+1) dimensional gapless Dirac semimetal. A closely related model is given by Gross-Neveu-Yukawa (GNY) model defined by the Lagrangian:
\bea
	\mathcal{L} &=& \sum_i \psi_i^\dag (\partial_\tau - i \vec \sigma \cdot \nabla) \psi_i + \frac12[ (\partial \varphi)^2 + (\nabla \varphi)^2 + m^2 \varphi^2]   \nn \\
	 &&~+ \frac{\lambda}{\sqrt{N}} \varphi \sum_i \psi^\dag_i \sigma^z \psi_i + \frac1{4!}\frac{u}{N} \varphi^4,
\eea
where the summation range is $i=1,...,N$, $\varphi$ is a real boson field similar to $\phi$ in Eq. (\ref{GN}), and $\lambda$, $u$ characterize the coupling strength. $m$ is the tuning parameter of the transition.

The GNY model describes the same universality class as the GN model in (2+1) dimensions. Different from GN model, the GNY model is renormalizable in four dimensions rather than in two dimensions \cite{justin2003, wilson1973}. However, here we consider (2+1) dimensions and use large-$N$ as a control parameter. Parallel to the situation in the GN model, the boson field $\varphi$ receives large renormalization from the bubble diagram shown in Fig. \ref{polarization} at large-$N$ limit: $\Pi_\varphi(i \omega_n, \vec k)=- \frac{\lambda^2}8 \sqrt{\omega_n^2 + k^2}$. The renormalized boson field has scaling dimension $[\varphi]=1$, such that the terms $(\partial \varphi)^2 + (\nabla \varphi)^2$ and $\varphi^4$ are irrelevant. Keeping the relevant degrees of freedom, and integrating out the boson field, we have
\bea
	\mathcal{L} &=& \sum_i \psi_i^\dag (\partial_\tau - i \vec \sigma \cdot \nabla) \psi_i - \frac{1}{2N} \frac{\lambda^2}{m^2} \Big( \sum_i \psi^\dag_i \sigma^z \psi_i \Big)^2.
\eea
It is identical to the GN model provided $ \frac{g}{\sqrt{2}}= \frac\lambda{m}$. Thus, in (2+1) dimensions, these two models are equivalent as far as long wavelength and low energy physics are concerned \cite{justin2003, justin1991, shen1991}. Because temperature is the only energy scale in the critical theory, the Lyapunov exponent should be a universal quantity. As a result, both GN model and GNY model would have same Lyapunov exponent, i.e., our calculation applies to the GNY model as well.

Besides the chiral Ising symmetry class discussed in above, GNY universality also includes other symmetry class, e.g., chiral XY and chiral Heisenberg universality classes \cite{justin2003, yu1993}. What about the Lyapunov exponents in these classes? In the large-$N$ calculation, the fermion flavor is much larger than bosons, such that boson propagator is dominated by the polarizations operator. In all these chiral classes, Dirac fermions are gapped out by the condensate of the boson modes. As a result, the polarization operator---$\Pi_{ij} \sim \int_k \text{Tr}[G \Gamma_i G \Gamma_j]$, where $\Gamma_i$ is the Yukawa coupling---are same as the chiral Ising class. Thus, we expect that all the chiral classes have the same Lyapunov exponent to the lowest order in $1/N$ expansion.


Intensive interests in exploring strongly interacting properties of the Dirac fermions in (2+1) dimensions \cite{herbut2006, son2007, honerkamp2008, qi2008, herbut2009a, herbut2009b, lahde2009, neto2012, polikarpov2013, vafek2014, wehling2014, herbut2015, li2015, sorella2016, li2017, jian2017} has been triggered by the experiments on graphene or gaphene-like materials \cite{firsov2005, geim2009} and cold atom systems loaded in optical lattices \cite{esslinger2013, esslinger2015, greiner2017} . These materials provide ideal platforms to realize spinless and/or spinful relativistic Dirac fermions and correspondingly, the GN criticality. The Lyapunov exponents in the GN criticality could hopefully be measured through OTOC in controllable quantum devices.

\begin{center}
{\bf ACKNOWLEDGEMENT} 
\end{center}
This work is supported in part by the NSFC under Grant No. 11474175 (S.-K.J. and H.Y.) and by the MOST of China under Grant No. 2016YFA0301001 (H.Y.).

\begin{widetext}

\renewcommand{\theequation}{A\arabic{equation}}
\setcounter{equation}{0}
\renewcommand{\thefigure}{A\arabic{figure}}
\setcounter{figure}{0}
\renewcommand{\thetable}{A\arabic{table}}
\setcounter{table}{0}
\section*{Appendix}

\subsection{The transition points in the Gross-Neveu model}
Here, we review the transition described by the GN model \cite{justin2003}. After Hubbard-Stratonovich transformation, the Lagrangian is quadratic in  fermion operators, and we can trace them out,
\bea
	\frac{S}{N}=- \text{Tr} \log(\partial_\tau - i \sigma \cdot \nabla+ m \sigma^z)+  \frac{m^2}{g},
\eea
where $m \equiv \frac{\phi}{\sqrt N}$. Assuming a uniform $m$, the saddle point equation is given by $\Big(\frac1g- \Omega(m) \Big) m =0$, where $\Omega(m) \equiv \int_k \frac1{\omega_n^2+ \epsilon_k^2}$ and $\epsilon_k= k^2+ m^2$. If $ \frac1g> {\Omega(0)}$, there is only one solution, $m=0$, and the model describes a symmetric phase. If $ \frac1g< {\Omega(0)}$, there are two solutions, $m=0$ and $g= \frac1{\Omega(m)}$. The $\Omega(m)$ can be calculated directly,
\bea
	\Omega(m) =  \int \frac{d^2 k}{(2\pi)^2} T\sum_n \frac1{\omega_n^2+ \epsilon_k^2}=  \int \frac{d^2 k}{(2\pi)^2} \frac{1}{2\epsilon_k} \tanh \frac{\epsilon_k}{2T} \approx \frac\Lambda{4\pi}- \frac{T}{2\pi} \log\cosh \frac{m}{2T}-  \frac{\log2}{2\pi} T.
\eea
where $\Lambda$ is the integration cutoff. Thus we have
\bea
	\frac1{g_c}- \frac1g= \frac1{2\pi\beta} \log\cosh \frac{\beta m}2 +  \frac1{2\pi\beta} \log2,
\eea
where the quantum critical point is given by $g_c= \frac{4\pi}\Lambda$. And the finite temperature transition is $\frac1{g_c}- \frac1{g(T)}= \frac{\log2}{2\pi} T$, as shown in the main text.

\subsection{The polarization functions}
At finite-temperature, the polarization function is given by
\bea
	\Pi(i\nu_n, \vec p) &=& - \int_k \text{Tr} [G(i\omega_n-i\nu_n , \vec k-\vec p) \sigma^z G(i\omega_n , \vec k) \sigma^z]\\
	&=& - \sum_{a,b} \int_k K_{ab}(\vec k, \vec k- \vec p) G_a(i\omega_n-i\nu_n , \vec k-\vec p) G_b(i\omega_n , \vec k) \\
	&=& - \sum_{a,b} \int_{\vec k} K_{ab}(\vec k, \vec k- \vec p) \frac{n(a|\vec k-\vec p|)-n(b|\vec k|)}{i\nu_n+ a|\vec k-\vec p|- b|\vec k|}.
\eea
where $K_{ab}(\vec k, \vec p)= \frac{1- ab \hat k\cdot \hat p}2$. Analytically continuing to real frequency, we have
\bea
	\Pi_R(\nu, \vec p) &=& \sum_{a,b} \int_{\vec k} K_{ab}(\vec k, \vec k- \vec p) \frac{n(b|\vec k|)-n(a|\vec k-\vec p|)}{\nu+ a|\vec k-\vec p|- b|\vec k|+ i\delta } \\
	&=& \sum_{a,b} \int_0^\infty \frac{dk}{2\pi} \int_{|k-p|}^{k+p} \frac{dq}{2\pi} \Big( \frac{p^2-(k-q)^2}{(k+q)^2-p^2} \Big)^{ab/2} \frac{n(b|\vec k|)-n(a|\vec k-\vec p|)}{\nu+ a|\vec k-\vec p|- b|\vec k|+ i\delta } \\
	&=& \int_p^\infty \frac{d\xi}{2\pi} \int_{0}^{p} \frac{d\eta}{2\pi} \Big[ \sqrt{ \frac{p^2-\eta^2}{\xi^2-p^2}} [n(\frac{\xi+\eta}2)-n(\frac{\xi-\eta}2)] \Big(\frac1{\nu-\eta+ i\delta }- \frac1{\nu+ \eta+ i\delta} \Big) \\
	&& + \sqrt{ \frac{\xi^2-p^2}{p^2-\eta^2}} [n(\frac{\eta+\xi}2)-n(\frac{\eta-\xi}2)] \Big(\frac1{\nu-\xi+ i\delta }- \frac1{\nu+ \xi+ i\delta} \Big) \Big],
\eea
where in the second equality we change the coordinate from $\theta$ to $q$, which are related by $\cos \theta= \frac{k^2+p^2-q^2 }{2kp}$, and in the third line we make the coordinate transformations, $\xi= k+ q, \eta= k- q$. Next, according to $\frac1{x \pm i\delta} = \mathcal{P} \frac1x \mp i \pi \delta(x)$, the dimensionless functions appearing in the main text are given by
\bea
	I_1(\Omega, P) &=& \frac{2}\pi \mathcal{P} \int_y^\infty d\xi \int_{0}^{y} d\eta \frac{ \sqrt{ \frac{P^2-\eta^2}{\xi^2-P^2}}\frac\eta{\eta^2-\Omega^2}\sinh \eta+ \sqrt{ \frac{\xi^2-P^2}{P^2-\eta^2}} \frac\xi{\xi^2-\Omega^2}\sinh \xi}{\cosh \xi+ \cosh \eta}- \int_0^\infty d\xi \tanh\xi , \\
		I_2(\Omega, P) &=& \begin{cases}
		\int_P^\infty  dz \sqrt{\frac{P^2-\Omega^2}{z^2-P^2}} \frac{\sinh \Omega}{\cosh \Omega + \cosh z}, \quad &\Omega<P, \\
		\int_0^P dz \sqrt{\frac{\Omega^2-P^2}{P^2-z^2}} \frac{\sinh \Omega}{\cosh \Omega + \cosh z}, \quad &\Omega>P,
		\end{cases}
\eea
where $\mathcal{P}$ means the principal value and $P>0, \Omega>0$. We have shifted the $\frac1{g_c} \equiv \int_0^\infty d\xi \tanh\xi$ to $I_i$ which necessarily cancels the divergent. In order to further simplify the functions, we change the argument $\Omega$ to $r\equiv \Omega/P$:
\bea
	I_1(P, r) &=& \frac{2P}\pi \int_1^\infty d\xi \int_{0}^{1} d\eta \frac{ \sqrt{ \frac{1-\eta^2}{\xi^2-1}}\frac\eta{\eta^2-r^2}\sinh P \eta+ \sqrt{ \frac{\xi^2-1}{1-\eta^2}} \frac\xi{\xi^2-r^2}\sinh P \xi}{\cosh P\xi+ \cosh P\eta}- \int_0^\infty d\xi \tanh\xi- \ln 2, \\
	I_2(P, r) &=& \begin{cases}
		  P \int_1^\infty dz \sqrt{\frac{1-r^2}{z^2-1}} \frac{\sinh Pr}{\cosh P r + \cosh P z}, \quad r<1, \\
		  P \int_1^\infty dz \sqrt{\frac{1-r^2}{z^2-1}} \frac{\sinh Pr}{\cosh P r + \cosh P z}, \quad r>1,
		  \end{cases}
\eea
By matching the result at zero temperature, we have added a constant $\ln 2$ to get the right answer. Now we divide the plane of $P-\Omega$ into four regions: region 1, $P\ll1, r<1$; region 2: $P\ll1, r>1$; region 1: $P\gg1, r<1$; region 1: $P\gg1, r>1$. In region 1 and region 2,
\bea
	I_1(P, r) &\approx& \frac{2P}\pi \int_1^\infty d\xi \int_{0}^{1} d\eta \frac{ \sqrt{ \frac{1-\eta^2}{\xi^2-1}}\frac\eta{\eta^2-r^2} P \eta+ \sqrt{ \frac{\xi^2-1}{1-\eta^2}} \frac\xi{\xi^2-r^2}\sinh P \xi}{\cosh P\xi+1}- \int_0^\infty d\xi \tanh\xi -\ln 2 \nn \\
	&\approx& \frac{2P^2}\pi  \int_1^\infty d\xi  \frac{1}{\sqrt{ \xi^2-1}}  \frac{1}{\cosh P\xi+1} \int_{0}^{1} d\eta\frac{\sqrt{1-\eta^2}\eta^2}{\eta^2-r^2}  \nn\\
	&& \quad\quad\quad + P \int_1^\infty d\xi   \Big[ \frac{\sqrt{ \xi^2-1}\xi}{\xi^2-r^2}\frac{\sinh P \xi}{\cosh P\xi+1}- \tanh P\xi\Big]- P \int_0^1 d\xi \tanh P\xi -\ln 2 \nn \\
	&\approx& \frac{4P^2}\pi  K_0(P) \int_{0}^{1} d\eta\frac{\sqrt{1-\eta^2}\eta^2}{\eta^2-r^2} - 2\ln 2+ O(P^2) \approx \begin{cases}
		P^2  K_0(P)(1-2r^2)- 2\ln 2, \quad &r<1, \\
		P^2 K_0(P)(1-2r^2+2r\sqrt{r^2-1})- 2\ln 2, \quad &r>1, \nn\\
	\end{cases} \\
	\\
	I_2(P, r) &\approx&  \begin{cases}
	P \sinh Pr \sqrt{1-r^2} \int_1^\infty dz \frac1{\sqrt{z^2-1}} \frac{1}{\cosh P z+1 } \approx 2K_0(P) P \sinh Pr \sqrt{1-r^2} , \quad &r<1 \\
	 P \sinh Pr \sqrt{r^2-1} \int_0^1 dz \frac1{2\sqrt{1-z^2}} = \frac\pi4 P \sinh Pr \sqrt{r^2-1} , \quad &r>1.
	\end{cases}
\eea
where $K_0$ denotes the Bessel function of the second type, and
in region 3 and region 4,
\bea
I_1(P, r) &\approx& \frac{2P}\pi \int_1^\infty d\xi \int_{0}^{1} d\eta \Big( \sqrt{ \frac{1-\eta^2}{\xi^2-1}}\frac\eta{\eta^2-r^2} \frac{1}{e^{P (\xi- \eta)}+ 1} + \sqrt{ \frac{\xi^2-1}{1-\eta^2}} \frac\xi{\xi^2-r^2} \frac{1}{e^{P ( \eta-\xi)}+ 1} \Big) - \int_0^\infty d\xi \tanh\xi- \ln 2 \nn \\
	&\approx& \frac{2P}\pi \int_1^\infty d\xi  \frac{e^{-P \xi}}{\sqrt{\xi^2-1}} \int_{0}^{1} d\eta \frac{\sqrt{1-\eta^2}\eta}{\eta^2-r^2}   + P \int_1^\infty d\xi \Big(\frac{\sqrt{\xi^2-1}\xi}{\xi^2-r^2} -\tanh P\xi \Big) - P\int_0^1 d\xi \tanh P\xi -\ln 2\nn \\
	&\approx& \begin{cases}
	\frac{2P}\pi K_0(P) \Big(-1- \frac12 \sqrt{1-r^2} \log\frac{2-r^2-2\sqrt{1-r^2}}{r^2} \Big)- \frac\pi2 P\sqrt{1-r^2}, \quad & r<1, \\
	\frac{2P}\pi K_0(P) (\sqrt{r^2-1} \text{arccsc} r-1), \quad & r>1,
	\end{cases} \\
I_2(P, r)  &\approx& \begin{cases}
 P \tanh Pr e^{Pr} \sqrt{1-r^2} \int_1^\infty dz \frac{e^{-Pz}}{\sqrt{z^2-1}} \approx P \tanh Pr \exp Pr K_0(P) \sqrt{1-r^2} , \quad &r<1, \\
 \frac\pi2 P\sqrt{r^2-1}, \quad &r>1.
\end{cases}
\eea
We summarize the functions as the dimensionless polarization function, $\Pi_R(\Omega, P)= I_1(\Omega, P)+ i I_2(\Omega, P)$:
\bea
	\Pi_R(\Omega, P) = \begin{cases}
					-2\ln 2+ K_0(P)(P^2- 2 \Omega^2) + i 2K_0(P) \sinh \Omega \sqrt{P^2- \Omega^2}, \quad &\Omega< P, P \ll 1, \\
					\\
					-2\ln 2- K_0(P) ( \sqrt{\Omega^2- P^2}- \Omega)^2 + i \frac{\pi}{4} \sinh \Omega \sqrt{\Omega^2-P^2}, \quad &\Omega> P, P \ll 1,\\
					\nn \\
					 -\frac{\pi}2 \sqrt{P^2- \Omega^2}+ \frac{2}{\pi} K_0(P)(-P- \sqrt{P^2- \Omega^2} \ln \frac{P-\sqrt{P^2- \Omega^2}}{\Omega})+ i \tanh\Omega e^{\Omega} K_0(P) \sqrt{P^2-\Omega^2}, \quad& \Omega< P, P \gg1, \\
					\\
					\frac{2}{\pi} K_0(P)(-P+ \sqrt{\Omega^2-P^2} \text{arccsc}\frac{\Omega}{P}) + i \frac{\pi}2 \sqrt{\Omega^2- P^2},\quad &\Omega>P, P \gg1.
	\end{cases}
\eea

\subsection{The quantum scattering rate}
To get the results in the low energy region, i.e., $P\ll 1$, we divide $\Gamma_P$ into two parts, corresponding to $b=\pm$. The first part gives rise to
\bea	
\Gamma_{P}^{(1)} &\approx& \int_P^1 \frac{KdK}{\cosh K} \int^{K+P}_{K-P} \frac{dQ}{2\pi} \frac1P  \sqrt{\frac{K^2-(P-Q)^2}{(P+Q)^2-K^2}} \frac{\frac{\pi}{4} \sqrt{(P+Q)^2-K^2}}{[-\ln 2- K_0(K) ( \sqrt{(P+Q)^2-K^2}- (P+Q))^2]^2} \nn
\\
&&+  \int_1^\infty \frac{KdK}{\cosh K \sinh K} \int^{K+P}_{|K-P|} \frac{dQ}{2\pi} \frac1P \sqrt{\frac{K^2-(P-Q)^2}{(P+Q)^2-K^2}} \frac{\frac{\pi}2 \sqrt{(P+Q)^2- K^2}}{ [\frac{2}{\pi} K_0(K)(-K+ \sqrt{(P+Q)^2- K^2} \text{arccsc}\frac{P+Q}{K}) ]^2}, \nn \\
&\approx&  \int_0^1 \frac{KdK}{\cosh K} \frac{\pi/4 }{[\ln 2+ K_0(K) K^2]^2} \int^{K+P}_{K-P} \frac{dQ}{2\pi} \frac{\sqrt{K^2-(P-Q)^2}}P \nn\\
&&+\int_1^\infty \frac{KdK}{\cosh K \sinh K} \frac{\pi/2 }{ [\frac{2}{\pi} K_0(K) K]^2} \int^{K+P}_{|K-P|} \frac{dQ}{2\pi} \frac{\sqrt{K^2-(P-Q)^2}}P \nn\\
&\approx& \sqrt{P} \times \Big[ \int_0^1 \frac{KdK}{\cosh K} \frac{\sqrt{K/8} }{[\ln 2+ K_0(K) K^2]^2} + \int_1^\infty \frac{KdK}{\cosh K \sinh K} \frac{\sqrt{K/2} }{ [\frac{2}{\pi} K_0(K) K]^2} \Big] ,
\eea
and the second part gives rise to
\bea
	\Gamma_{p}^{(2)} &\approx& \int_P^1  \frac{KdK}{\cosh K} \int^{K+P}_{K-P} \frac{dQ}{2\pi} \frac1P \sqrt{\frac{(P+Q)^2-K^2}{K^2-(P-Q)^2}} \frac{2K_0(K) \sqrt{K^2- (P-Q)^2}}{[-\ln 2+ K_0(K)(K^2- 2 (P-Q)^2)]^2}  \\
	&&+   \int_1^\infty  \frac{e^{K}KdK}{\cosh^2 K} \int^{K+P}_{K-P} \frac{dQ}{2\pi} \frac1P \sqrt{\frac{(P+Q)^2-K^2}{K^2-(P-Q)^2}} \frac{K_0(K) \sqrt{K^2-(P-Q)^2}}{[\frac{2}{\pi} K_0(K)K]^2} \\
	&\approx&   \int_P^1  \frac{KdK}{\cosh K} \frac{2K_0(K)}{[\ln 2+ K_0(K)K^2]^2} \int^{K+P}_{K-P} \frac{dQ}{2\pi} \frac{\sqrt{(P+Q)^2-K^2}}P   \\
	&&   +\int_1^\infty  \frac{e^{K}dK}{\cosh^2 K} \frac{1 }{(\frac{2}{\pi})^2 K_0(K)K} \int^{K+P}_{K-P} \frac{dQ}{2\pi} \frac{\sqrt{(P+Q)^2-K^2}}P  \\
	&\approx& \sqrt{P}\times \Big[ \int_0^1  \frac{KdK}{\cosh K} \frac{2\sqrt{2K}K_0(K)/\pi}{[\ln 2+ K_0(K)K^2]^2} + \int_1^\infty  \frac{e^{K}dK}{\cosh^2 K} \frac{\pi \sqrt{K/8}}{ K_0(K)K}\Big].
\eea
Thus, in the low energy limit, we have $\Gamma_P \propto \sqrt{P}$.

\subsection{The kernel function}
Assuming the eigenfunction of the largest eigenvalue is rotationally invariant, we have
\bea
	(-i \nu+ 2 \Gamma_{k, a}) f_a(\nu, k)= \frac1N  \sum_b \int \frac{k'dk'}{2\pi} [\mathcal{M}_{ab}^{(1)}( k, k')+\mathcal{M}_{ab}^{(2)}(k, k')] f_b(\nu, k'),
\eea
where one-rung kernel is given by
\bea
	\mathcal{M}_{ab}^{(1)}(k, k')= \int \frac{d\theta}{2\pi} K_{ab}(\vec k, \vec k')D_{W}(a|\vec k|-b|\vec k'|, q)= \frac1{kk'} \int \frac{qdq}{2\pi} \Big( \frac{q^2-(k-k')^2}{(k+k')^2-q^2}\Big)^{ab/2}D_{W}(a|\vec k|-b|\vec k'|, q) ,
\eea
and the two-block kernel is given by
\bea
	\mathcal{M}_{ab}^{(2)}(k, k') =\int \frac{d\omega''}{2\pi}\frac{k''dk''}{2\pi} \frac1{kk'} \sqrt{\frac{k^{\prime\prime2}-\omega^{\prime\prime2}}{(2k-a\omega'')^2-k^{\prime\prime2}}} \sqrt{\frac{k^{\prime\prime2}-\omega^{\prime\prime2}}{(2k'-b\omega'')^2-k^{\prime\prime2}}}  \frac{|D_{R}(\omega'', k'')|^2}{\cosh \frac{\beta(k-a\omega'')}2 \cosh \frac{\beta(k'-b\omega'')}2}, \label{M2A}
\eea
where we have used
\bea
	&& \int \frac{d\theta'}{2\pi} K_{bb'}(\vec k', \vec k'- \vec k'')
	= \frac1{k'} \int_{|k'-k''|}^{k'+k''} \frac{dq}{2\pi} \Big( \frac{(q+k')^2-k^{\prime\prime2}}{k^{\prime\prime2}-(q-k')^2} \Big)^{-bb'/2},
\eea
and
\bea
	&& \sum_{b'} \frac1{2k'} \int_{|k'-k''|}^{k'+k''} \frac{dq}{2\pi} \Big( \frac{(q+k')^2-k^{\prime\prime2}}{k^{\prime\prime2}-(q-k')^2} \Big)^{-bb'/2}2\pi \delta(b|\vec k'|-\omega''-b' q) =  \frac1{k'} \sqrt{\frac{k^{\prime\prime2}-\omega^{\prime\prime2}}{(2k'-b\omega'')^2-k^{\prime\prime2}}}, \label{frequency}
\eea
and similarly for integration over angle $\theta''$. Note that in Eq. (\ref{frequency}), the of integration range of $q$, $(|k'-k''|, k'+k'')$, will affect the integration range of frequency $\omega''$ because of the delta function. 
After simplifications, the one-rung kernels are
\bea
	\mathcal{M}_{\pm}^{(1)}(k, k') = \frac1{kk'} \int_{|k-k'|}^{k+k'} \frac{qdq}{2\pi} \sqrt{ \frac{q^2-(k\mp k')^2}{(k\pm k')^2-q^2}} D_{W}(k \mp k', q) , \nn \\
\eea
and the two-rung kernels are
\bea
	\mathcal{M}_{+}^{(2)}(k, k') = \frac1{kk'} \int \frac{k''dk''}{2\pi} \Big(\int_{-k''}^{\text{min}(k-|k''-k|,k'-|k''-k'|)} + \int_{\text{max}(k+|k''-k|,k'+|k''-k'|)}^{\text{min}(k''+2k,k''+2k')} \Big)\frac{d\omega''}{2\pi} \nn \\
	\times \sqrt{\frac{k^{\prime\prime2}-\omega^{\prime\prime2}}{(2k-\omega'')^2-k^{\prime\prime2}}} \sqrt{\frac{k^{\prime\prime2}-\omega^{\prime\prime2}}{(2k'-\omega'')^2-k^{\prime\prime2}}}  \frac{|D_{R}(\omega'', k'')|^2}{\cosh \frac{k-\omega''}{2T} \cosh \frac{k'-\omega''}{2T}}, \label{M2+} \\
	\mathcal{M}_{-}^{(2)}(k, k') = \frac1{kk'} \int \frac{k''dk''}{2\pi} \Big(\int_{-k''}^{\text{min}(k-|k''-k|, -k'-|k''-k'|)} + \int^{k''}_{\text{max}(k+|k''-k|, |k''-k'|-k')}  \Big)\frac{d\omega''}{2\pi} \nn \\
	\times \sqrt{\frac{k^{\prime\prime2}-\omega^{\prime\prime2}}{(2k-\omega'')^2-k^{\prime\prime2}}} \sqrt{\frac{k^{\prime\prime2}-\omega^{\prime\prime2}}{(2k'+\omega'')^2-k^{\prime\prime2}}}  \frac{|D_{R}(\omega'', k'')|^2}{\cosh \frac{k-\omega''}{2T} \cosh \frac{k'+\omega''}{2T}}.\label{M2-}
\eea

\end{widetext}

\end{document}